\documentclass[pra,aps,multicols,epsfig,twocolumn]{revtex4}

\newcommand\be{\begin{eqnarray}}
\newcommand\ee{\end{eqnarray}}
\newcommand\ba{\begin{array}}
\newcommand\ea{\end{array}}
\def\r{\rangle}
\def\l{\langle}

\def\cV{{\cal V}}

\begin{document}

\title{Towards quantum-based privacy and voting}
\author{
Mark Hillery$^{1}$,
M\'ario Ziman$^{2,3}$,
Vladim\'\i r Bu\v zek$^{2,4,}$\footnote{E-mail address: buzek@savba.sk}, and
Martina Bielikov\'a$^{5}$
}
\affiliation{
$^{1}$~Department of
Physics, Hunter College of CUNY, 695 Park Avenue,  New York, NY 10021, USA
}
\affiliation{
$^2$~Research Center for Quantum Information, Slovak Academy of Sciences,
D\'ubravsk\'a cesta 9, 845 11 Bratislava, Slovakia
}
\affiliation{
$^3$~Faculty of Informatics, Masaryk University, Botanick\'a 68a, 602 00 Brno, Czech Republic
}
\affiliation{
$^{4}$~Abteilung f\"{u}r Quantenphysik, Universit\"{a}t Ulm, 89069 Ulm, Germany
}
\affiliation{
$^5$~{\rm Quniverse}, L\'\i\v s\v sie \'udolie 116, 841 04 Bratislava, Slovakia
}

\date{20 August  2005}

\begin{abstract}
The privacy of communicating participants is often of paramount importance, but in some situations
it is an essential condition. A typical example is a fair (secret) voting. We analyze in detail
communication privacy based on quantum resources, and we propose new quantum protocols.
Possible generalizations that would lead to voting schemes are discussed.
\end{abstract}

\maketitle

Every day people have to make important decisions that should remain secret.
Protecting the privacy of those decisions, if their results are to be communicated, can
be a challenging problem.  In this
Letter we will consider a special instance of multi-party decision making.
Consider a group of people  who have to make
a common decision,
i.e. choosing one of several possible (prescribed) options. In many cases the
fairest (democratic) way of making the decision is to
{\it vote}.  Reliable voting protocols should satisfy a number of conditions \cite{schneier},
three of which are:
{\it i)} security,
{\it ii)} verifiability, and {\it iii)} privacy.
The security condition guarantees that all users can influence the
result only by casting a {\it single} valid vote. That is, each
voter can vote
just once (non-reusability), only legitimate users can vote
(eligibility) and no one can learn any intermediate result (fairness).
The strongest version of the verifiability requirement is that each voter can verify
the correctness of the result, however none of the voters is able to prove
how he or she voted. This prevents vote buying.

A voting scheme satisfying all  properties except the privacy condition
is easy to implement. Privacy is related to
the secrecy of the ballots, or equivalently to the anonymity of the
voters. Ideally, no one should be able to tell how any of the voters voted.
Such multi-party communication protocol
is known as {\it secret}, or {\it anonymous voting}.

In the voting process the initial information $I_i$ created by voters (their votes) is
transformed into the final outcome corresponding to the information
$I_f$.
Usually $I_i$ is much larger than $I_f$. The voting can be formalized
as mapping $\cV:X\times \dots\times X\to Y$, where $X,Y$ represent sets
of individual voting options and a set of final results,
respectively. The input/output information
can be expressed in terms of the cardinalities of the sets $X,Y$, i.e.
$I_i=N \log_2|X|$ ($N$ is the number of participants) and
$I_f=\log_2|Y|$.
For example, $N$ parliamentarians during a voting procedure
create $I_{i}=N\log_2 3$ bits of information, given the choices: $X=\{accept,refuse,abstain\}$.
However, the final result can be represented only by $I_f=\log_2 3$
bits with $Y=\{acceptance,refusal,undecided\}$.

Hiding the identity of the voters seems to be very difficult to
achieve, because the information
can often be traced back to its origin. In a public
election the collected paper ballots are mixed in a ballot box, which
could ensure the anonymity of a voter. However, the
ballots could be marked in such a way that it is possible
to identify the voters. Thus perfect
privacy no longer holds. The first protocol to guarantee voting privacy (see Ref.~\cite{chaummix})
was based on the so-called {\it MIX net machines}.
Since then several secret voting protocols
based on cryptographic primitives such as anonymous broadcast
\cite{chaummix} or blind signatures  \cite{bos,pfitzmann},
have been proposed.
Some of the properties of these protocols are even unconditionally
secure, and others are guaranteed in the computational sense, i.e.
they are based on one-way functions. David Chaum \cite{chaum}
suggested a solution of the so-called {\it dinning cryptographers
  problem} that can be used to implement a secret voting protocol.
Let us briefly describe this procedure, which
guarantees unconditionally the anonymity of the voters.

Three cryptographers are having a dinner in their favorite
restaurant. After ordering their food, the waiter comes and informs
them that someone has already paid for the dinner. The problem is to determine whether
it has been paid for by one of them, or by someone else (for instance, by the NSA).
The cryptographer who paid (if indeed one of dinning cryptographers took care of the bill)
wants to remain anonymous. The three cryptographers
can resolve the problem by using the following protocol: Each
pair $\{ j,k\}; (j,k=1,2,3)$
of cryptographers toss a coin, i.e. they generate a random bit $c_{jk}$. The third cryptographer
cannot see the result.
Each cryptographer announces the logical (mod2) sum of the two bits he shares, i.e., the
cryptographer $k$ announces the value $s_k=\sum_{j\ne k} c_{jk}$, unless he is the one who
paid for the dinner. The one who paid the bill announces the opposite value,
i.e. $s_k=1+\sum_{j\ne k}c_{jk}$. If the
dinner was paid by the NSA then $\sum_k s_k = 0$. If not, then $\sum_k
s_k=1$. The identity of the potential payer remains completely secret.

This protocol can be easily extended to an arbitrary number of users,
and a slightly modified version can be used for voting.
Each pair of voters shares a random integer
(key) $|c_{jk}|\le N$, where $c_{jk}=-c_{kj}$.
Each voter chooses either $v_k=0$ (``no'') or $v_k=1$  (``yes'').
He broadcasts the message $s_k =
v_k+\sum_{j\ne k} c_{jk}$. Because of $c_{jk}=-c_{kj}$ it is valid that
$\sum_k s_k=\sum_k v_k$.  Finally, each of the users can compute
the sum and find out the total number of the ``yes'' votes. Let us note
that all operations are modulo $N$. Privacy in this scheme is assured,
but the protocol is not secure and cheating is easy.
One cannot guarantee that voters will not ``vote'' an arbitrary number times
$v_k\le N$, i.e. the result can be easily manipulated. However, there
exists a modification that solves this problem and provides security
based on the RSA protocol (for more details see Refs.~\cite{bos,chaum}).

We would now like to consider whether the paradigm and tools of quantum (information) theory
and in particular, the quantum cryptography \cite{gisin}
can help us implement tasks related to secret voting and maintaining the anonymity of the voters.

The first analysis of identity protection based on quantum
protocols was performed  by Christandl and Wehner (see
Ref.~\cite{christandl}), who used generalized GHZ states to {\it
anonymously broadcast} not only classical bits, but also qubits.
 Recently, two approaches (see Refs.~\cite{martina} and
\cite{vaccaro}) to {\it quantum voting} have been independently proposed.
In Ref.~\cite{vaccaro} Vaccaro et {\it al.} have proposed a scheme for
quantum voting, in which the quantum protocol is used to
ensure the voter's privacy.  In their scheme, the voters and an authority share an entangled
state with a fixed number of particles, $N$.  This state is a superposition of states of the
form $|N-n,n\rangle$, where the the first number is the number of particles possessed by
the authority and the second is the number of particles possessed by the voters.  The
number of yes votes in encoded in the relative phases of the states making up the
superposition.  In Ref.~\cite{martina} quantum
approaches towards privacy and voting based on entangled qudit states have been discussed.
The quantum-voting protocol introduced there is the subject of this paper.
Both these protocols utilize similar resources, i.e. the {\it ballot} (voting)
state in both cases is a maximally entangled state, and each voter registers his
vote by a local operation. On the other hand, the two approaches suggest different
implementations of quantum voting.  Both will need further development.

A closely related problem of {\it
anonymous oblivious transfer} has been studied in
Ref.~\cite{mquade}. The privacy problem is relatively new even in
classical information theory. It seems (at least intuitively)
that quantum systems may be more suitable and more efficient in
achieving this goal.

Let us start with the analysis of the privacy and the voting problem in
the framework of quantum theory. Our aim is to use
quantum information to ensure privacy.
In principle, one can imagine two general schemes for a voting procedure:
{\it i)} the distributed-ballots scheme (DB) and {\it ii)} the
travelling-ballot scheme (TB).
In the DB a voter obtains his own ballot, he or she performs
voting operation and sends the ballot back.
The TB is a scheme in which a single ballot/container is
travelling (is sent) between voters and everyone performs the voting operation on
the same physical system. Both, the DB and the TB scenarios can be formalized in
the same way. Physically quantum voting is performed by transformations of
some quantum system. Let us denote by $|\Omega_0\rangle$ the initial state of
the system (the quantum ballot) and by $U^{(j)}_{k}$ a transformation performed by the $j$th
user voting for the option $k\in X$. After the voting has concluded, the ballot is sent to the
authority, who performs
a measurement $M$ on the ballot. The outcome of the measurement
 $r$ is associated with the result of the voting, $r\in Y$. Let us denote by $\vec{v}=(k_1,\dots,k_n)$
the particular collection of votes and by ${\cal V}(\vec{v})\in Y$ the
result of the process of voting.  After the
voting is completed the system is described by the state
$|\Omega_{\vec{v}}\rangle =U^{(n)}_{k_n}\dots U^{(1)}_{k_1} |\Omega_0 \rangle$.
The difference between these two schemes is that in the DB scheme we work with
a composite system of $N$ particles, and the operations for different users
mutually commute.

In what follows we will assume that the initial state $|\Omega_0\rangle$ is pure
and voting operations are represented by unitary maps.
The privacy of votes is reflected by the following set of conditions
\be
\label{pc}
|\l \Omega_{\vec{v}_1}|\Omega_{\vec{v}_2}\r|=0,\ {\rm iff} \ \
{\cal V}(\vec{v}_1)\ne {\cal V}(\vec{v}_2)\; ; \\
|\l \Omega_{\vec{v}_1}|\Omega_{\vec{v}_2}\r|=1,\ {\rm iff}\ \
{\cal V}(\vec{v}_1)= {\cal V}(\vec{v}_2)\; .
\nonumber
\ee
These conditions guarantee  that finally only $I_f$ bits of information are
available and the identity of the voters is securely hidden. Our task is to find collections of voting operations
$\{U^{(j)}_k\}$ and an initial state $|\Omega_0\rangle$ such that the above
conditions are satisfied. In what follows we will simplify the task by
assuming that each participant uses the same collection of operations,
i.e. $U_k^{(j)}=U_k^{(j^\prime)}\equiv U_k$ for all $j,j^\prime$ and
all $k$.

Let us start by analyzing the simplest case of two voters. In this case the privacy property
does not make much sense, because after a public announcement of the result
each voter  can deduce how the other
participant has voted. However, it can be of interest
to some third party, particulary in the case of an {\it undecided} result,
i.e. when votes do not coincide. Let us consider the TB scheme
first. The set of possible results restricts from below the dimension of the
required quantum system, i.e. $\dim{\cal H}\geq |{\cal Y}|$.
In our case $Y=\{acceptance,refusal,undecided\}$ and
consequently at least a qutrit is needed to perform the voting.
Let us assume  the state $|\Omega_0\r=|0\r$ and voting transformations
$U_{no}=I$ and $U_{yes}=U$, where $U$ is defined via transformations
$|0\r\to|1\r\to|2\r\to|0\r$. It is easy to verify that
Eqs.~(\ref{pc}) are fulfilled, however this system does not guarantee the
secrecy of votes. In the proposed protocol it is very easy to learn the actual
state of the voting, i.e. the intermediate result. One way to avoid
this problem is to use an {\it authority} who prepares the initial
state of the ballot and finally reads the result.  We shall now show how this can protect
voter privacy.

In what follows we present
a protocol with an {\it honest} (non-cheating) authority that utilizes entangled states:
The authority
prepares two qutrits that serve as a ballot in a maximally entangled state
$|\Omega_0\r=\frac{1}{\sqrt{3}}(|00\r+|11\r+|22\r)$.
One of the qutrits is sent to the voters. Using the voting operation
$U_{yes}$ they produce mutually orthogonal states associated with
mutually exclusive results of voting. In this way one can guarantee that
intermediate measurements performed individually by voters who might want to learn
an intermediate voting result provide no information about how the voting
has progressed.  The state of the particle seen by the voters is, at all times,
 simply a total mixture. This
protocol can be directly generalized to $N$ participants by choosing the
ballot state to be $|\Omega_0\r=\frac{1}{\sqrt{N}}\sum_{k=0}^{N-1} |k\r |k\r$ and having
$U_{yes}|k\r = |k+1\r$.  There is a possibility of voter collusion in this scheme.  If two voters
want to find out how the voters between them voted (how many total yes votes these
voters made) then the first voter can measure the second particle of the ballot state when
he gets it, placing it in one of the states $|k\r$ and disentangling it from the first particle, and then
the second colluding voter can again measure second ballot particle when he receives it.  By
comparing their results, they can determine the number of yes votes cast by the people who
voted between them.  There is a rather high cost to this type of cheating, however; the result
of the voting as determined by the authority will be completely random.

Next we shall again consider the case of two voters, but
we shall analyze the DB scheme. Let us start with the assumption that each
of the participants obtains a ballot represented by a single qubit.
In the previous paragraph we have argued
that the system has to be at least three-dimensional. Although the
system of two qubits is four-dimensional, the DB scheme allows the users
to make only local voting operations, i.e. single-qubit unitary
transformations. The question is, whether two qubits are sufficient to implement secret voting.
Suppose that $U_0,U_1$ and $V_0,V_1$ are voting operations of the
voters named $U$ and $V$, respectively. Let us denote by $|\Omega_0\r$
the initial state of the two ballot qubits. The privacy conditions (\ref{pc})
yield  the following system of equations
\be
\label{r1}
0&=&\l\Omega_0|U_0^\dagger U_1\otimes I|\Omega_0\r=\l\Omega_0|I\otimes V_0^\dagger V_1|\Omega_0\r\; ;\\
\label{r2}
1&=&\l\Omega_0|U_0^\dagger U_1\otimes V_1^\dagger V_0|\Omega_0\r\; ;\\
\label{r3}
0&=&\l\Omega_0|U_0^\dagger U_1\otimes V_0^\dagger V_1|\Omega_0\; ;\\
\label{r4}
0&=&\l\Omega_0|U_1^\dagger U_0\otimes I|\Omega_0\r=\l\Omega_0|I\otimes V_1^\dagger V_0|\Omega_0\r\; .
\ee
Due to the fact that $U_0^\dagger U_1$ and $V_0^\dagger V_1$ are unitary single-qubit
transformations, they can be expressed as
$U_0^\dagger U_1=I\cos\nu+i(\vec{m}\cdot\vec{\sigma})\sin\nu$
and $V_0^\dagger V_1=I\cos\theta+i(\vec{n}\cdot\vec{\sigma})\sin\theta$.
It also holds that $(U_0^\dagger U_1)^\dagger=U_1^\dagger U_0$
and $(V_0^\dagger V_1)^\dagger=V_1^\dagger V_0$.
>From Eqs.~(\ref{r1}) and (\ref{r2}) it follows that
\be
\cos\nu\cos\theta+\sin\nu\sin\theta\l\Omega_0|\vec{n}\cdot\vec{\sigma}\otimes\vec{m}\cdot\vec{\sigma}|\Omega_0\r=1\, ,
\ee
which together with Eq.~(\ref{r3}) results in the condition $2\cos\nu\cos\theta=1$. However,
combining Eqs.(\ref{r1}) and (\ref{r4}) we find that
$\cos\theta=0$ and $\cos\nu=0$. This is a contradiction and therefore
two-dimensional systems are not sufficient to implement the DB scheme.
One needs at least two qutrits. Note that in the DB scheme the
fairness requirement holds, i.e. nobody can learn intermediate
results.

To proceed further let us recall the following property.
If the protocol $(|\Omega_0\r,\{U_k\})$ satisfies the privacy
conditions, then so does the protocol
$(|\Omega_0^\prime\r,\{U_k^\prime\})$, where $U_k^\prime=U_k V$ and
$|\Omega_0^\prime\r=(V^{\otimes N})^\dagger|\Omega_0\r$. This
means we can always choose $U_{0}=I$, i.e.
without the loss of generality we can assume that $U_{no}=I$
and $U_{yes}=U$. Applying the privacy conditions from Eq.~(\ref{pc})
we obtain equations
\be
\l\Omega_0|U\otimes U|\Omega_0\r&=&0 \; ;
\nonumber \\
\l\Omega_0| U\otimes I|\Omega_0\r&=& 0\; ;
\nonumber \\
 \l\Omega_0|I\otimes U|\Omega_0\r&=& 0\; ;
\nonumber
 \\
\l\Omega_0|U^\dagger\otimes U|\Omega_0\r&=&1\; .
\ee
Our task is to find
a solution of this set of equations. Because $U$ is unitary,
its eigenvalues are
of the form $e^{i\eta_j}$ and
$U=\sum_j e^{i\eta_j}|j\rangle\langle j |$, where $|j\rangle$ is
the eigenvector corresponding to the eigenvalue  $e^{i\eta_j}$.
Let us consider the state
$|\Omega_0\r=\sum_j |\alpha_j|^2 |j\r\otimes |j\r$,
where, again, $\{|j\r\}$ are
eigenvectors of $U$. Using this Ansatz the above identities gives us
the following equations $\sum_j |\alpha_j|^2 e^{i2\eta_j}=0$,
$\sum_j |\alpha_j|^2 e^{i\eta_j}=0$, and $\sum_j |\alpha_j|^2=1$.
The last of these equations is just the normalization condition for the
state $|\Omega_0\r$, and is satisfied.
We have already shown that a qubit is not a sufficient resource
for quantum voting and
larger dimensional systems have to be sent to both voters. Assuming
$d=3$ we find that the equations for parameters $\eta_j,\alpha_j$ have multiple
solutions. Here, however,  we
will restrict ourselves to $|\alpha_j|=1/\sqrt{d}=1/\sqrt{3}$.
The possible solutions then form a one-parameter set and among them
is the following one
\be
\nonumber
U &=& e^{i2\pi/3}|0\r\l 0|+e^{i4\pi/3}|1\r\l 1|+e^{i6\pi/3}|2\r\l 2|\; ;\\
|\Omega_0\r&=& \frac{1}{\sqrt{3}}(|0\r |0\r+|1\r |1\r+|2\r |2\r)\; .
\ee
This solution
can be easily extended to an arbitrary number of participants.
Let us suppose that there are $N$ voters and the authority that
distributes one qudit ($d>N$) to each voter. In accordance with
the case of two voters let the authority prepare and distribute the state
$|\Omega_0\r=\frac{1}{\sqrt{d}}\sum_j |j\r^{\otimes N}$. The voters
will apply one of the two operations: either $U_{no}=I$, or $U_{yes}=U=\sum_k
e^{ik2\pi/d}|k\r\l k|$, depending on their decision. After performing the
voting operations they send the qubits back to the authority, who now
possesses the state $|\Omega_m\r=U^{\otimes m}\otimes
I^{\otimes (N-m)}|\Omega_0\r=\sum_k e^{imk2\pi/d}|k\r^{\otimes N}$,
where $m$ is the number of voters who choose to vote ``yes'', i.e. the
result of the voting. Note that this state contains no
information about the particular voter, only the total number of
``yes'' votes is recorded. This guarantees the  privacy of the
individual votes.
The distinguishability of different outcomes (different numbers
of ``yes'' votes) is
guaranteed by the orthogonality condition
i.e. $\l\Omega_m|\Omega_{m^\prime}\r=\delta_{mm^\prime}$.

This protocol protects the identity of the voters from a curious,
but not malicious,
authority and from other voters.  We assume that the authority follows
the protocol,
but does whatever he can beyond this to determine the individual
votes.
In this case,
because the state he receives contains no information about
who voted how, he can do nothing.  If, after the voting,
a voter intercepts a particle from another voter, he will be able to learn
nothing, because any  subset of particles has
a reduced density matrix
proportional to the identity.

The protocol can also be used in the dinning-cryptographers problem.
It has further uses
as well.  A simple generalization of the protocol can be used also for
what was called in Ref.~\cite{vaccaro} an anonymous survey.
Imagine that $N$ people want to determine the
total amount of money they have, but each individual does not want
reveal how much money
he or she has. These people use the same method,
except that each person votes ``yes''
a number of times corresponding to the number of Euros (s)he has.
In the resulting state,
$|\Omega_m\r$, $m$ will be equal to the total number of Euros, but the
contributions of individual partners will be unknown.
This is an example of quantum secure
function evaluation \cite{klauck,crepeau}.

In a sense this voting protocol can be thought of as a
generalization of the classical scheme. However, like its
classical counterpart, it does not
completely possess the properties of the security and the verifiability.
In fact, we have focused our attention mainly on the un-traceability
of voters ( privacy).
To achieve a completely secure voting scheme the protocol has to be
improved.  One of the security loopholes
is the possibility for voters to vote more than
once.  This type of cheating, however, is not
guaranteed to produce the desired result, because the states $|\Omega_m\r$
count only the number of ``yes'' votes modulo $d$. If too many voters vote
 ``yes'' too many times, the number $m$ can become larger than $d$, and because
$|\Omega_m\r$ records the number of ``yes'' votes only modulo $d$, the final
recorded number $m$ could be small. In fact, even
if there is a subset of cooperating
voters it would be difficult to know what the effect of
such cheating would be, and the
voting process would become a strategic game.

In order to prevent voters from registering more than one vote (a
complete non-reusability of ballots) we can proceed as follows:
Besides the qudit from the state $|\Omega_0\r$ each of the
participants receives two additional ``voting'' qudits, one in the
state $|\psi(\theta_y)\r$ and the second one in the state
$|\psi(\theta_n)\r$. These two qudits represent the ``yes" and the
``no'' votes, respectively. Both of the states are of the form
$|\psi(\theta)\r=\frac{1}{\sqrt{d}}\sum_k e^{ik\theta}|k\r$.
The angle $\theta_{y}$ is equal to $(2l_{y}\pi /d)+\delta$ and
$\theta_{n}$ is equal to $(2l_{n}\pi /d)+\delta $, where $l_{y}$
and $l_{n}$ are integers between $0$ and $d$ and $\delta$ is an
angle between $0$ and $2\pi /d$.  In addition, we assume that
$(l_{y}-l_{n})N <d$, where, as before, $N$ is the number of
voters.  This condition is necessary in order that different
voting results be distinguishable. The integers $l_{y}$ and
$l_{n}$ and the angle $\delta$ are not known to the voters.
Depending on his (her) choice the voter combines either
$|\psi(\theta_y)\r$, or $|\psi(\theta_n)\r$, with the original
ballot particle, i.e. creates a system composed from the ballot
and the voting qudits. Then (s)he performs a two-qudit measurement
that is specified by a set of projectors $P_r=\sum_j |j+r\r_{b}\l
j+r|\otimes |j\r_{v}\l j|$, where the subscript $b$ denotes the
ballot qudit while the subscript $v$ denotes the voting qudit.
Registering the outcome $r$ the voter applies the operation
$V_{r}=I_b\otimes \sum_j |j+r\r_v\l j|$ to the  voting qudit and
sends both (the ballot and the voting) qudits back to the
authority.  The remaining unused qudit must be kept, or destroyed
in order to secure the privacy of the registered vote. Let us see
how this works in a bit more detail for the first voter.
Assuming the vote ``yes'' we obtain
$P_r(\frac{1}{\sqrt{d}}\sum_k
|k\r|\psi(\theta_y)\r|k\r^{\otimes(N-1)})= \frac{1}{d}\sum_k
e^{i(k-r)\theta_y}|k\r|k-r\r|k\r^{\otimes(N-1)}$. After the
``correcting'' unitary operation $V_r$ the $N+1$ qudits are in the
state $|\Omega_1\r=\frac{1}{\sqrt{d}}e^{-ir\theta_y}\sum_k
e^{ik\theta_y} |k\r^{\otimes (N+1)}$, where the global phase $e^{-ir\theta_y}$ is
irrelevant. If $m_y=m$ the voters voted ``yes'' and $m_n=N-m$
voters voted ``no", the authority gets back the state
$|\Omega_{m}\r=\frac{1}{\sqrt{d}}\sum_k
e^{ik(m_y\theta_y+m_n\theta_n)}|k\r^{\otimes 2N}$.  Here the
phase factor can be rewritten as follows
$e^{ik(m_y\theta_y+m_n\theta_n)}=e^{ikm\Delta} e^{ikN\theta_n}$,
where $\Delta=\theta_y-\theta_n = 2\pi (l_{y}-l_{n})/d$.  Because
the states $\sum_k e^{ipk2\pi /d}|k\r^{\otimes 2N}$ are orthogonal
for different values of $p$, for $p$ an integer between $0$ and
$d-1$, we see that from the state $|\Omega_m\r$ the authority can
determine the value of $p$ corresponding to
this state, which is just $m(l_{y}-l_{n})$.  This allows
him to determine $m$, because he knows both $l_{y}$ and $l_{n}$.  Note that $p$ should
always be a multiple of $l_{y}-l_{n}$ if the voters are using their proper ballot states.  If after
 measuring the ballot state, the authority finds a value of $p$ that is not a multiple of $l_{y}-l_{n}$,
then he knows that someone has cheated.  A voter who wants to vote more than once is
faced with the problem of determining what $\theta_{y}$ or $\theta_{n}$ is, and this cannot
be done from just a single state.  In fact, what the voter is faced with is a problem of phase
estimation, and the best he can do is to determine the phase to within an accuracy of
order $1/d$.  If he makes an error (illegal operation), the value of $\Delta$ will not, in general be a multiple of
$2\pi /d$, and this will cause the authority to obtain different results for $m$ if the voting
procedure is repeated several times.  The choice $l_y - l_n =\pm 1$ maximizes the chance
of different results being found if cheating has occurred, and thereby makes it most likely that
the cheating will be detected. We also note that the no-cloning theorem makes it
impossible for a voter to simply copy the voting states.  Summarizing, we see that the
casting of multiple votes can be prevented by the use of voting states and repeating the
voting procedure several times in order to make sure that the result is the same each time.
If the results vary, cheating is taking place, and the results are discarded.

If the authority is malicious as well as curious, one of the new
security problems that arises is that of  the {\it state identification},
i.e. verifying whether the initial ballot state is correct, and
whether the voting states provided by the authority are also correct.
The authority could cheat either by sending a ballot state that is a
 product state, so that each voter's particle is
decoupled from all of the others, or by sending voting states that are
different for each voter. In principle, if the authority is required
to supply a large number of both kinds of states to the voters, they
can verify the correctness of the states  by performing
state tomography.   A less drastic test is for the voters to take subsets of ballot particles
or subsets of voting particles corresponding to the same choice (all yes or all no) and to
measure them.  For example, a set of "yes" voting states can be measured in order 
to determine whether the state is completely symmetric or not.  If it is not,
then the authority is trying to cheat by sending different voting states to different voters.
Another possibility, at least in regard to the ballot state, is for the voters
to prepare that state by themselves. However, this is second option is
somewhat restrictive, because it requires that participants (voters)
have to meet as the same place (or they need quantum resources
for a remote state preparation).  This problem could perhaps be addressed by having
two authorities, one who prepares the states and one who counts the votes.  Here one would
have to assume that they are not both dishonest and cooperating with each other.  Clearly
much work remains to be done.

In summary, we have proposed quantum protocols
that guarantee the anonymity of participants in voting procedures and can
be used in several complex communication tasks.
The advantage of quantum voting protocols is that
security that is based on the laws of quantum mechanics rather than assumptions about
computational complexity.
Our approach to quantum voting is based on voters applying local operations to
an entangled state, a feature it shares with
the approach proposed by Vaccaro {\it et al.} \cite{vaccaro}.
As we have noted, the approaches discussed here are based on entangled states of qudits,
while in \cite{vaccaro} particle number entanglement is used.  This would suggest that the
schemes would lend themselves to different physical implementations.  Within our approach, we
have presented a method of preventing voters from cheating, that is from casting more
than one vote.  Hence, for the attacks discussed here, this scheme can meet two requirements
 for a voting system, privacy and security.  The are still many
questions that remain open and deserve further detailed investigation. Primarily, it is
the analysis of the security of quantum voting against more sophisticated attacks, such as
collaborating parties, the use of illegal voting operations, and cheating authorities that deserve
further attention.  In addition it would be useful to find physical implementations for some of
these proposals for small numbers of voters that might enable a prototype system to be
constructed.

We thank Jan Bouda for his help with the
literature on classical privacy and voting schemes.
This work was supported by the National
Science Foundation under grant PHY 0139692,
by the European
Union  projects QGATES, QUPRODIS and CONQUEST,  by the Slovak
Academy of Sciences via the project CE-PI, by the project
APVT-99-012304, and by the Alexander von Humboldt Foundation.


\end{document}